\documentstyle[12pt]{article}
\title{
Low-Energy Effective Action of N=2 Gauge Multiplet
Induced by Hypermultiplet Matter
}

\author{A.T. Banin$^{1}$, I.L. Buchbinder$^{2}$\footnote{joseph@tspu.edu.ru, joseph@fma.if.usp.br}
, N.G. Pletnev$^{1}$\footnote{pletnev@math.nsc.ru}}

\date{{\it
${^{1}}$Institute of Mathematics, \\
Novosibirsk, 630090, Russia,\\
\vspace{0.7cm}
${^{2}}$Instituto de F\'isica, Universidade de S\~ao Paulo, \\
P.O. Box 66318, 05315-970, S\~ao Paulo, Brazil \\
and \\
Department of Theoretical
Physics, \\ Tomsk State Pedagogical University, \\ Tomsk, 634041,
Russia } }

\begin{document}

\begin{titlepage}
\maketitle
\begin{abstract}
We study one-loop effective action of hypermultiplet theory coupled to
external N=2 vector multiplet. We formulate this theory in N=1 superspace
and develop a general approach to constructing derivative expansion of the
effective action based on an operator symbol technique adapted to N=1
supersymmetric field models. The approach under consideration allows to
investigate on a unique ground a general structure of effective
action and obtain both N=2 superconformal invariant (non-holomorphic)
corrections and anomaly (holomorphic) corrections. The leading low-energy
contributions to effective action are found in explicit form. 
\end{abstract}
\thispagestyle{empty}
\end{titlepage}

%-----------------------------------------------
\newcommand{\be}{\begin{equation}}
\newcommand{\ee}{\end{equation}}
\newcommand{\bea}{\begin{eqnarray}}
\newcommand{\eea}{\end{eqnarray}}
\newcommand{\pc}{\hbar}

%-----------------------------------------------

\section{Introduction}
Effective action containing the quantum corrections to classical action
plays in quantum theory a role analogous to one of action functional in
classical theory. Being found, the effective action allows to investigate
a broad spectrum of quantum properties associated with off-shell
behavior. Therefore it is not wonder that the effective action is one of
the central objects of quantum field theory.

In field models possessing some global or gauge symmetries on a classical
level the exact effective action contains full information concerning
these symmetries in quantum theory or their violation. Presence of the
symmetries imposes the rigid restrictions on a structure of the effective
action and allows sometimes to fix it very significantly in terms of
proper functionals invariant under the symmetries. The bright examples of
such models are the extended supersymmetric field theories.

One of the main approaches to practical evaluating the effective action is
momentum (derivative) expansion where the effective action is investigated
in form of a series in derivatives of its functional arguments. Keeping
the lowest terms of such an expansion leads to a notion of low-energy
effective action which can be described by local effective lagrangian.
Another approach is the known loop expansion where the leading (one-loop)
contribution to effective action is given by functional determinant of
some (pseudo)differential operator. Both these approaches are used very
often together.

The paper under consideration is devoted to study a structure of
low-energy effective action in hypermultiplet model coupled to external
abelian N=2 vector multiplet. The various aspects of effective action in
field models possessing N=2 supersymmetry attracted recently very much
attention due to famous work by Seiberg and Witten \cite{sw} where exact
low-energy effective action has been found in N=2 super Yang-Mills theory
with gauge group SU(2) spontaneously broken down to U(1). It is turned out
that just extended supersymmetry was one of the essential points allowing
to establish general non-perturbative structure of the effective
action. The result by Seiberg and Witten has later been generalized for
various gauge groups and coupling to a matter (see \cite{hph} for modern review).
Another remarkable result was obtained by Dine and Seiberg \cite{ds} and
concerned an exact structure of part of low-energy effective action
depending on N=2 superfield strengths in N=4 super Yang-Mills theory with
gauge group SU(2) spontaneously broken down to its abelian subgroup. Such
a theory can be treated as a specific N=2 supersymmetric model and its
extended supersymmetry has played a crucial role in obtaining the exact
form of low-energy effective action.

A remarkable feature of supersymmetric field models consists in the fact
that the low-energy effective action can be written as a sum of two
contributions. One of them is integral over full superspace and another
one is integral over chiral subspace of general superspace (plus
conjugate). Therefore the low-energy effective action in such models is
described by two types of effective lagrangians: chiral and general or
holomorphic and non-holomorphic. We point out that a possibility of
holomorphic corrections for N=1 SUSY models was firstly demonstrated in
papers \cite{wjj} and for N=2 SUSY models in papers \cite{vmn}. Non-holomorphic
superfield effective lagrangian was constructed in \cite{bky} (see also general
discussion in \cite{idea}).

We concentrate the attention on the N=2 SUSY models containing
an interaction with vector multiplet. In this case a part of effective
action depending only on a vector multiplet fields is written in the form
\be\label{n2}
\Gamma [{\cal W}]=
\left(\int d^8z\;{\cal F}({\cal W})
+ h.c. \right) +
\int d^{12}z \; {\cal H}({\cal W},\bar{\cal W})+\ldots
\ee
where ${\cal W}$ is N=2 superfield strength \cite{gsw}, $z$ are the N=2
superspace coordinates, $d^{8}z$ and $d^{12}z$ mean the chiral
and general N=2 superspace measures. ${\cal F}({\cal W})$ is
called holomorphic effective potential and ${\cal
H}({\cal W},{\bar{\cal W}})$ is called non-holomorphic effective
potential. The dots mean the terms depending on covariant
derivatives of the strengths. In arbitrary N=2 SUSY models the
holomorphic effective action ${\cal F}({\cal W})$ determines low
-energy behavior and ${\cal H}({\cal W},\bar{\cal W})$
corresponds to next to leading corrections (see f.e. \cite{henn}).
However in the theories possessing quantum superconformal
invariance , for example in N=4 super Yang-Mills theory,
holomorphic effective action is trivial (it is proportional to
${\cal W}^{2}$) and namely non-holomorphic effective potential
forms leading contribution to low-energy effective action. Recent
progress in finding the holomorphic and non-holomorphic
contributions to effective action in various N=2,4 SUSY models
is discussed in refs \cite{wgr}-\cite{bk}. In particular, a manifestly N=2
supersymmetric approach to calculating holomorphic effective
action was developed in refs \cite{bbb} on the base of concept of
harmonic superspace \cite{Gikos}. This approach includes also a general
N=2 superfield background field method \cite{bbk}, \cite{bko}. Structure of
leading low-energy contributions to effective action of N=4 SYM
theory has been investigated in recent refs. \cite{pu}, \cite{ct}, \cite{bp}. It is worth
to point out a significance of detailed study of N=4 SYM
effective action for understanding classical supergravity/quantum
gauge theory duality \cite{mtheor}, \cite{Tse}.

At present, the holomorphic effective potential is well established.
However a structure of non-holomorphic effective potential for arbitrary
N=2 models is still unclear. The known solid result corresponds to the
contribution of the form $({\cal W}{\bar{\cal W}})^{2}$ (see
first paper in refs \cite{bbb}). Situation changes drastically
in N=4 SYM theory or in the models possessing quantum N=2
superconformal symmetry. Here due to this symmetry one can get
one-loop effective action for constant field background in terms
of so called N=2 superconformal invariants \cite{bk}.  However the
approach \cite{bk} can not be applied literally to arbitrary N=2 models
since they are not N=2 superconformal invariant on quantum level.

The paper under consideration is just devoted to developing  a general
method for evaluating low-energy one-loop effective action in arbitrary
N=2 supersymmetric models. Our purpose consists in construction of the
derivative expansion of the effective action preserving N=1 supersymmetry
and gauge invariance. Another N=1 supersymmetry in nonmanifest but as we
will see the final result for leading contribution in constant field
background can be interpreted in N=2 SUSY terms.

We evaluate the effective action in the hypermultiplet model coupled to
external abelian N=2 vector multiplet using a realization of the model in
terms of N=1 superspace. This model is simple enough and allows to
illustrate all basic steps of general derivative expansion technique
discussed earlier (see various implementations of this technique in \cite{we}).

For calculating one-loop effective action of corresponding N=1
superfield theory we develop a general approach based on a technique of
the operator symbols closely connected with mathematical theory of
deformation quantization (see the references in subsection 3.2). A main
dignity of such an approach is a possibility to reformulate a problem of
evaluating the traces of the operators acting in superspace as problem of
calculating some integrals of suitable superfields with specific
non-commutative multiplication rule containing all quantum aspects of the
initial problem. We demonstrate that this approach is very efficient for
constructing derivative expansion of one-loop superfield effective action.

The paper is organized as follows. In Section \ref{action} we describe the properties
of the models and discuss a formal definition of the effective action.
Section \ref{EA} is devoted to general structure of the effective action, a
overview of a method of operator symbols which we apply for
evaluating the effective action and the specific features of
implementations of this methods to N=1 superfield theories.
In Section \ref{kernels} we carry out the calculations of the
low-energy effective action for constant
field background
\bea\label{backgr}
{\cal W}| =\Phi=Const, \: D^{i}_{\alpha}{\cal W}|=\lambda^{i}_{\alpha}=Const,
& & \nonumber \\
D^{i}_{(\alpha}D_{\beta )i}{\cal W}|=f_{\alpha\beta}=Const, \:
D^{\alpha (i}D^{j)}_{\alpha}{\cal W}|=0. & &
\eea
and obtain a general result including both (known earlier) holomorphic
and non-holomorphic effective potentials within a single method. Section 5
is devoted to a summary of the results and the prospects. Appendices
contain some details of the calculations.

\section{Description of the Model}\label{action}

We consider the hypermultiplet model interacting with external abelian
N=2 vector multiplet. Our purpose is to integrate over hypermultiplet
fields and construct an effective action depending on the vector
multiplet fields.

As well known the model under consideration can be formulated by
different (on-shell equivalent) ways: in terms of component fields, in
terms of N=1 superfields, in terms of (constrained) N=2 superfields
and in terms of unconstrained harmonic and projective superfields. In our case for
constructing the effective action we use the simplest realization of
the hypermultiplet in terms of N=1 chiral superfields. Although such
a realization does not preserve manifest N=2 supersymmetry it allows
to apply an efficient and well developed technique of N=1 superfield
quantum field theory (see f.e. \cite{idea}).

The action of the model in above realization of the hypermultiplet is
written as a sum of external fields action $S_{0}$ and hypermultiplet
action coupled to the external fields $S$.
\bea\label{act}
S_{0}&=&{1 \over 4g^2}[\int d^6z\,
{1 \over 2}W^{\alpha}W_{\alpha} + \int d^8z\, \bar{\Phi}{\rm e}^{-V}\Phi
{\rm e}^{V}], \\
S &=&
 \int d^8z \,(\bar{Q}_{+}{\rm e}^{V}Q_{+}+Q_{-}
{\rm e}^{-V}\bar{Q}_{-}) +
i\int d^6z \, Q_{-}\Phi Q_{+} +
i\int d^6 \, \bar{z} \bar{Q}_{+}\bar{\Phi}\bar{Q}_{-}.\label{int}
\eea
Here $Q_{+}$ and $Q_{-}$ are two N=1 chiral superfields
with opposite $U(1)$ charges; $V$ and $\Phi$ are N=1 vector multiplet
superfield and chiral superfield respectively, together they form
N=2 vector multiplet and the gauge coupling is included in the field definitions.
The actions $S_{0}$ and $S$ are N=1 supersymmetric by
construction. However they are invariant under hidden extra N=1 supersymmetry
$$
\delta\Phi = \epsilon^{\alpha}W_{\alpha}, \quad
\delta\bar{\Phi}=\bar{\epsilon}^{\dot{\alpha}}\bar{W}_{\dot{\alpha}},
\quad \delta\nabla=\epsilon\Phi, \quad
\delta\bar{\nabla}=-\bar{\epsilon}\bar{\Phi},
$$
$$
\delta W_{\alpha}=-\epsilon_{\alpha}\bar{\nabla}^2\bar{\Phi} +
i\bar{\epsilon}^{\dot{\alpha}}\nabla_{\alpha\dot{\alpha}}\Phi,
\quad \delta \bar{W}_{\dot{\alpha}}=-\bar{\epsilon}_{\dot{\alpha}}
\nabla^2\Phi +
i\epsilon^{\alpha}\nabla_{\alpha\dot{\alpha}}\bar{\Phi}.
$$
These transformations form together with manifest N=1 supersymmetry
transformations the full set of N=2 supersymmetry
transformations. Here $W_{\alpha}$ is the strength corresponding to
N=1 gauge superfield $V$. Besides, we have introduced the covariant
derivatives in background field vector representation.
$$
\nabla_{\alpha}= {\rm e}^{-{V \over 2}}D_{\alpha}
{\rm e}^{V \over 2}, \quad \bar{\nabla}_{\dot{\alpha}}=
{\rm e}^{V \over 2}\bar{D}_{\dot{\alpha}} {\rm e}^{- {V \over 2}},
$$
which satisfy usual constrains
$$
\{\nabla_{\alpha},\nabla_{\dot{\alpha}}\}=i\nabla_{\alpha\dot{\alpha}},
\quad \{\nabla_{\dot{\alpha}}, \nabla_{\beta\dot{\beta}} \}=
\epsilon_{\dot{\alpha}\dot{\beta}}W_{\beta}, \ldots
$$
and covariant chiral superfields
$$
\Phi_{c}= {\rm e}^{V \over 2}\Phi {\rm e}^{-{V \over 2}}, \quad
\bar{\Phi}_{c}= {\rm e}^{-{V \over 2}}\bar{\Phi}{\rm e}^{V \over
2},
$$
subject to the constrains
$\bar{\nabla}\Phi_{c}=\nabla\bar{\Phi}_{c}=0$.
The details of background vector representation see in
\cite{ggrs}, \cite{idea}.

The superfields $W$, $\bar{W}$ and $\Phi$, $\bar{\Phi}$ are the N=1
projections of N=2 gauge strengths ${\cal W}$ and $\bar{\cal W}$
$$
{\cal W}=\Phi+\eta^{\alpha}W_{\alpha}-\eta^2\bar{\nabla}^2\bar{\Phi}+
{i \over 2}\eta^{\alpha}\bar{\eta}^{\dot{\alpha}}\nabla_{\alpha\dot{\alpha}}\Phi
+ {i \over 2}\eta^2\bar{\eta}^{\dot{\alpha}}\nabla_{\alpha\dot{\alpha}}W^{\alpha}
+ {1 \over 4}\eta^2\bar{\eta}^2\Box\Phi,
$$
$$
\bar{\cal W}=\bar{\Phi}+\bar{\eta}^{\dot{\alpha}}\bar{W}_{\dot{\alpha}}-
\bar{\eta}^2\nabla^2\Phi+ {i \over 2}\eta^{\alpha}\bar{\eta}^{\dot{\alpha}}\nabla_{\alpha\dot{\alpha}}\bar{\Phi}
+ {i \over 2}\bar{\eta}^2\eta^{\alpha}\nabla_{\alpha\dot{\alpha}}
\bar{W}^{\dot{\alpha}} +
{1 \over 4}\eta^2\bar{\eta}^2\Box\bar{\Phi}.
$$
These relations allow to link the forms of N=2 supersymmetric
functionals written in terms N=1 and N=2 superfields.

It is worth to point out that the model under consideration is
not only N=2 supersymmetric but it possesses two more classical
symmetries. First, it is gauge invariant (see f.e. \cite{ggrs}) and second,
it is N=2 superconformal invariant, the corresponding superconformal
transformations are given in \cite{kt}

\section{The EA and Derivative Expansion Method}\label{EA}

\subsection{General Definition of Effective Action}

We introduce the effective action $\Gamma$ of the model under
consideration by the standard way
\be\label{1}
{\rm e}^{i\Gamma} = \int{\cal D}Q_{+}{\cal
D }Q_{-}\,{\rm e}^{i(S_{0}+S)}
\ee

Since the action S is quadratic in (quantum) hypermultiplet superfields
$Q_{+}$ and $Q_{-}$ the effective action $\Gamma$ has the
following structure
\be\label{2}
\Gamma = S_{0}+\Gamma_{(1)}
\ee
where quantum correction $\Gamma_{(1)}$ to classical action $S_{0}$
of the gauge multiplet induced by hypermultiplet is formally
written as follows
\be\label{init}
\Gamma_{(1)}=-{i \over 2}\ln {\rm Det}(\hat{H}) =
-{i \over 2} {\rm Tr}\ln \hat{H},
\ee
Here $\hat{H}$ is some differential operator associated with action S and acting
in space of N=1 chiral and antichiral superfields $Q_{+}$ and $Q_{-}$. Its
explicit form will be presented bellow. Eq (\ref{init}) expresses the formal path
integral (\ref{1}) in terms of formal functional determinant. To provide a sense
to these formal relations we have to give an informal definition allowing
to compute unambiguously the functional determinants of the differential
operators acting in N=1 superspace.

As we already pointed out in Section \ref{action} the model under consideration
possesses by three classical symmetries: N=2 supersymmetry, gauge symmetry
and N=2 superconformal symmetry. Supersymmetry and gauge invariance are
not anomalous on quantum level but superconformal symmetry is expected to
be broken down due to (one-loop) divergences containing some scale.
Therefore the quantum correction $\Gamma_{(1)}$ is N=2
supersymmetric and gauge invariant functional and hence it have
to depend only on N=2 strengths ${\cal W}$ and $\bar{\cal W}$ or
on their N=1 projections $W_{\alpha}, \bar{W}_{\dot \alpha},
\Phi$ and $\bar{\Phi}$.

Taking into account an appearence of the superconformal anomaly one can
write $\Gamma_{(1)}$ in the form
\be\label{4}
\Gamma_{(1)}=\Gamma_{(1)}^{0}+\Gamma^{1}_{(1)}
\ee
where $\Gamma_{(1)}^{0}$ ia a functional generating superconformal
anomaly (its N=2 superconformal variation is equal to anomalous
current) and $\Gamma_{(1)}^{0}$ is superconformal invariant
functional. Of course, decomposition (\ref{4}) is not unique since one
can add an arbitrary superconformal functional to
$\Gamma_{(1)}^{0}$ and it still will generate the given
superconformal anomaly. The main purpose of this paper is
developing a technique for efficient evaluation of the functional
$\Gamma_{(1)}$. We show that the decomposition (\ref{4}) arises quite
naturally in our approach and a role of the $\Gamma_{(1)}^{0}$ is
played by the known Seiberg's type holomorphic effective action.

Since the effective action is expressed in form of functional determinant
of the (superfield) differential operator $\hat{H}$ its calculation can be carried
out on the base of Fock-Schwinger proper-time technique appropriately
formulated in superspace (see the aspects of such a formulation in
ref. \cite{bky}, \cite{idea}, \cite{we})

The evaluation of the determinants of the (pseudo)differential operators
always involves some kind of regularization. For actual computations of
the effective action we use elegant $\zeta$-function
regularization formulated directly in superspace. Within this
regularization scheme the functional determinant
$
{\rm Det}_{\zeta}(\hat{H})=\exp (-\zeta'_{\hat{H}}(0))
$
is supersymmetric and gauge invariant. As a result the effective action
looks like
$
\Gamma_{(1)}=-{i \over 2}\zeta '_{\hat{H}} (0)
$
and $\zeta$-function is defined by the expression
\be\label{zeta}
\zeta_{\hat{H}} (s)= {1 \over \Gamma (s)}\int_{0}^{\infty}dT\,T^{s-1}{\rm Tr}
({\rm e}^{-{T\over \mu^2}\hat{H}}) =
{1 \over \Gamma (s)}\int_{0}^{\infty}dT\,T^{s-1}
\int dz \,K({T\over \mu^2}),
\ee
where $dz$ is an appropriate (super)space measure and $\mu$
is a renormalization point, which is introduced to make $T$
dimensionless. One can show that the dependence on
the parameter $\mu$ occurs only in those terms that correspond
to divergences for other renormalization schemes
(proper-time cut-off, dimensional regularization, etc.).
The quantity $K(T)$ is the coincidence limit of heat kernel which
can be represented in form of Schwinger-DeWitt expansion over
proper time $T$
\be\label{deck}
K(T)= {1 \over (4\pi T)^2}\sum_{f=0}^{\infty}a_{f}T^{f}.
\ee
Here $a_{f}$ are DeWitt-Seely coefficients which are the scalars
constructed from the coefficients of the operator $\hat{H}$.
The representation (\ref{deck}) is used to isolate the infinities
and finite contributions in effective action by Schwinger's method.
Diverging at small $T$ terms correspond to $\mu$-dependence
of the effective action. The series (\ref{deck}) automatically
leads to the following expansion of function $\zeta_{\hat{H}}(s)$
\be\label{decz}
\zeta_{\hat{H}}(s)= \sum_{f}a_{f}\zeta_{f}(s).
\ee
This asymptotic expansion encodes information about short-distance
behavior of the effective action in invariant terms.

The action $S$ can be written as
\be\label{intmat}
S=\int d^8z
\pmatrix
{Q_{-}{\rm e}^{-{ V\over 2}},& \bar{Q}_{+}{\rm e}^{V\over 2} \cr}
\pmatrix{
1 & \Phi_c {\nabla^2 \over \Box_{-}}\cr
\bar{\Phi}_c {\bar{\nabla}^2 \over \Box_{+}} & 1 \cr
}
\pmatrix
{{\rm e}^{-{ V\over 2}}\bar{Q}_{-} \cr {\rm e}^{ V\over 2}Q_{+}}
\ee
where $d^{8}z=d^{4}x d^{2}\theta d^{2}\bar{\theta}$ with
$z^{M}=(x^{m}, \theta_{\alpha}, \bar{\theta_{\dot{\alpha}}})$ be
N=1 superspace coordinates. Let us rewrite the operator $\hat{H}$
associated with eq (\ref{intmat}) in more convenient form. We would like
to avoid explicit dependence on background gauge prepotential $V$.
To do that we introduce covariantly chiral functional
variation
${\delta Q(z)\over\delta Q(z')}=\bar{\nabla}^2\delta(z-z')$,
and define the operator $\hat{H}$ as second variation of the
action (\ref{intmat}) according to previous relation. It leads to
the following manifestly N=1 supersymmetric and gauge invariant
form of the operator
\be\label{kinet}
\hat{H}=
\pmatrix{
\bar{\nabla}^2 \nabla^2 & \Phi_c \bar{\nabla}^2 \cr
\bar{\Phi}_c \nabla^2   & \nabla^2 \bar{\nabla}^2 \cr
}.
\ee
Equations (\ref{init}, \ref{zeta}, \ref{kinet}) are considered here as the
definition of the effective action $\Gamma$ given by formal path
integral (\ref{1}).  It is important to point out that this
definition does not appeal to calculating the path integral
(\ref{1}) via direct integration over unconstrained chiral
superfields $Q_{+}$ and $Q_{-}$ in order to obtain the standard
$1/p^{2}$ propagator (see f.e. \cite{hst}, \cite{gsz}). As well known
such a scheme requires to introduce an infinite tower of ghosts
which contributes to effective action. Our
definition of effective action by means of Eqs
(\ref{init},\ref{zeta},\ref{kinet}) avoids making use of everything
associated with these ghosts and looks like most simple from
computational point of view.

As was recently shown some known tricks of evaluating the
one-loop effective action based on factorization of the functional
determinants can be ambiguous because of so called multiplicative anomaly
(see f.e. \cite{Eliz}). In its essence the multiplicative
anomaly is a violation of the equalities ${\rm Det}(AB)={\rm
Det}(A){\rm Det}(B)={\rm Det}(B){\rm Det}(A)$ for functional
determinants of the formal infinite matrices. The matter is all
these determinants need regularization and there no guarantee that
above equalities always survive after regularization. Our
definition of the effective action on the base of Eqs
(\ref{init},\ref{zeta},\ref{kinet}) do not appeal to any factorization
triks and therefore allows to avoid in principle the problem of
possible multiplication anomalies.

\subsection{Star-Product Algebras of Function and Derivative Expansion of
Effective Action}\label{method}

The evaluation of the effective action is always based on the computing
traces of some operator functions. Exact calculations of such traces is
possible only for very specific cases when the eigenvalues and
eigenfunctions of the operator under consideration are known, that rather 
exception then a rule. 

Recently the proof of existence of a "formal trace" has been given in ref.
\cite{Fed} within a so called deformation quantization in the
form very similar to the Schwinger-DeWitt expansion
$$
{\rm Tr}(\hat{O})={1 \over n!\,\hbar^n}
\int_{X}d\mu\, (O\omega^n+\hbar\tau_{1}(O)+\hbar^2\tau_{2}(O)+\cdots),
$$
where $\tau_{k}$ are local expressions in $O$.
Here, under the deformation quantization we mean a formal deformation of 
commutative algebra of functions with $\hbar$ as the deformation parameter on arbitrary symplectic manifold to the noncommutative algebra of quantum 
observables subject to Dirac's correspondence principle only.
Calculating the traces via equations (\ref{1}, \ref{init}) closely related 
to the index theorem and naturally leads to star product algebras of functions 
(see recent ref. \cite{Stern}). Therefore, the natural language
for the trace handling is the star product algebras on functions.

A proposal to reformulate an analysis in the operator algebras on a
symbol calculus language copying the operator product has been
introduced by Berezin \cite{Ber}. The basic role in this
construction plays a concept of symbol $\sigma({\cal O})$ of the
operator $\hat{\cal O}$. The symbol is a classical function of finite
number of the variables $\gamma^{A}$ associated with the operator
$\hat{\cal O}(\gamma)$ ordered by certain manner
(we are implying $\sigma(\hat{\gamma}_{A})=\gamma_{A}$).

In order to set the stages for computations, let us briefly
review some basic notions used and terminology.
A symplectic manifold $M^{2n}$ can be treated as a cotangent
fiber bundle $X=(M^{2n}, M^{n}, T^{*}_{x}M^{n}, \omega)$ with
the base space $M^{n}$, fiber $T^{*}_{x}M^{n}$ and a
fundamental symplectic two-form $\omega$ in the form
$\omega={1\over 2}\omega_{AB}d\gamma^{A}\wedge d\gamma^{B}$.
where the local coordinates
$\gamma^{A}=(p_{i},x^{i})$, $\gamma\in M^{2n}$,
$x\in M^{n},\,p\in T^{*}_{x}M^{n}$.

In particular, nondegenerate matrix $\omega_{AB}$ is a constant
matrix in the Darboux coordinates and $\omega^{AB}$ defines
the standard Poisson bracket
$\{f,g\}_{\rm PB}=f\stackrel{\leftarrow}{\partial}_{A}
\omega^{AB}\stackrel{\rightarrow}{\partial}_{B}g$ which
is a noncommutative product in $M^{2n}$ in contrast to
a pointwise product. The dynamical behavior of the system
is then controlled by a function $H$ defined on a manifold
through the vector field associated with $\omega$ by means of
its differential.

The starting point of the symbol $\leftrightarrow$ operator correspondence is an
associative algebra that defines a
noncommutative space and can be described in terms of a set of
operators $\hat{\gamma}_{A}$ and relations
\be\label{alg}
[\hat{\gamma}_{A},\,\hat{\gamma}_{B}]=\hbar\omega_{AB}(\hat{\gamma}).
\ee
Among these relations the most known are:
1) canonical structure $\omega_{AB}= Const$;
2) Lie-algebra structure
$\omega_{AB}=\omega_{AB}^{C}\hat{\gamma}_{C}$;
3) quantum space structure
$\omega_{AB}=\omega_{AB}^{CD}\hat{\gamma}_{C}\hat{\gamma}_{D}$.
One considers the generators $\hat{\gamma}_{A}$ as coordinates
and let the algebraic structure is the formal power series
in these coordinates modulo to the relations (\ref{alg}).
It means that the power series whose elements reordered with the
help of these relations are considered as equivalent. The known
{\em Heisenberg-Weyl algebra}
$\{\hat{\gamma}_{A}\}=\{\hat{P}_{i},\hat{Q}^{j}\}$
is defined by the commutation relation
$[\hat{Q}^{j},\,\hat{P}_{i}]= i\hbar\delta_{i}^{j}$.

Let us introduce an operator family as a Fourier transform
$s$-parameterized by a weight function $w_{s}(u,v)$ of
displacement operators since they form a complete operator basis
\be\label{swks}
\hat{\Omega}(p,q;s)=\int du\,dv\;{\rm e}^{i(vq+up)}
w_{s}(u,v){\rm e}^{{i\over \hbar}(u\hat{P}+v\hat{Q})}.
\ee
Because of any
operator obeying certain conditions can be expanded
in terms of the complete operator basis
we can present an operator $A(\hat{P},\hat{Q})$
in the enveloping Heisenberg algebra in the following way
\be\label{rule}
A(\hat{P},\hat{Q}) =
\int {d^n p d^n q \over (2\pi)^n} \;
A_{-s}(p,q)\hat{\Omega}(p,q;s), \;
A_{s}(p,q)={\rm Tr}(\hat{A}\hat{\Omega}(p,q;s)),
\ee
where coefficient $A_{s}(p,q)$ is a smooth function on
$T^{*}M$ and is called $s$-symbol of the operator
$A(\hat{P},\hat{Q})$.

The construction mentioned above can be extended to a more general case.
Actually, the concept of the deformation quantization is related
to Weyl's quantization procedure (i.e. $s=0$). In this procedure
a classical observable $A(\gamma)$, some square integrable
function on phase space $X=(\gamma, \omega)$,  is one-to-one
associated to a bounded operator $\hat{A}$ in the Hilbert space
by the {\em Weyl mapping}
\be\label{wmapin}
\hat{A}= \int_{X}
d\mu(\gamma)\, A_{-s}(\gamma) \hat{\Omega}(\gamma;s).
\ee
An inverse formula which maps an operator into its symbol by {\em
Wigner mapping},  is given by the trace formula
\be\label{wignm}
A(\gamma)= {\rm Tr}(\hat{\Omega}(\gamma;-s)\hat{A}),
\ee
Both  formulae (\ref{wmapin}, \ref{wignm}) are determined by the
choice of the {\em Stratonowich-Weyl kernel}
$\hat{\Omega}(\gamma;s)$, which is the Hermitian operator
family parameterized by $s$ and constructed from
the operators $\hat{\gamma}_{A}$.
The Stratonowich-Weyl kernel (or a {\em quantizer} and also a
{\em dequantizer}) possesses
by a number of properties (see for example ref. \cite{prop}):
$\hat{\Omega}$ is injectiv; $\hat{\Omega}$ is self-adjoint;
unit trace ${\rm Tr}(\hat{\Omega})=1$;
covariance $U(g)\hat{\Omega}(x)U(g^{-1})=\hat{\Omega}(g\cdot x)$;
traciality ${\rm Tr}(\hat{\Omega}(x)\hat{\Omega}(y))=\delta(x,y)$.
One can see from expressions (\ref{swks}) and (\ref{rule}) that
in general a symbol can possess a parametric dependence on
$\hbar$ by formal power series.

Correspondence (\ref{wmapin}) relates
$\hat{A}$ to $A_{-s}(\gamma)$ via integration. In practical calculation
it is also helpful to employ a differential form of this relation
$$
\hat{A}=A_{-s}(-i\partial_{\gamma})\hat{\Omega}(\gamma;s)|_{\gamma=0}
$$
Various $s$ related to various {\em ordering prescriptions}
in the corresponding enveloping algebra. This
means that we can choose several different rules of normal
ordering for operator products. For instance, the {\em Weyl ordering} (totally symmetrized
operator product) is often a preferred choice for physical
applications since it treats self-adjoint operators
$\hat{P}, \hat{Q}$ symmetrically. This ordering prescription has
specific features leading to a possibility to construct the real symbols
for the operators (i.e. complex conjugation is an algebra
anti-automorphism). Other ordering prescriptions convenient for
practical calculations are the standard $PQ$ (all $\hat{P}$ are
disposed from the left of all $\hat{Q}$) and antistandard $QP$.

The behavior of physics system is described
in terms of states and observables. Both of them are represented by
a set of functions on some space $X$. The space $X$ is a set of
points with some particular structure. All the information
about $X$, without any loss, can be retrieved from the algebra of
the functions alone. Moreover, the existence of a such space $X$
even may not be necessary, if to transfer all relevant information
concerning a  physical theory into the algebra of functions.
This is well-known Gelfand-Naimark duality: i.e. every
structure defined on $X$ has a natural counterpart on the algebra of
functions. Particularly, canonical $\omega_{AB}$ is transferred into
the Poisson bracket on the algebra of functions in accordance
with the Dirac's correspondence principle.
The symbol of the non-commutative product of
operators can be written as a non-local star product
$$
(A\star B)(\gamma)\leftrightarrow\hat{A}\hat{B},
$$
which for a constant Poisson structure is called {\em Moyal product} and
might be treated for a particular case of Weyl - ordering
prescription in integral and differential forms as follows
\bea\label{hstar}
(A\star B)(p,q) &=&
\int {d\xi\,d\eta \over 2\pi\hbar}\,{d\xi'\,d\eta' \over 2\pi\hbar}\;
{\rm e}^{{i \over \hbar}S}
A(\xi,\eta)B(\xi',\eta'),\nonumber \\
(A\star B)(p,q)&= &A(p, q)
{\rm e}^{i\hbar(
{\stackrel{\leftarrow}{\partial} \over \partial q}
{\stackrel{\rightarrow}{\partial} \over \partial p}-
{\stackrel{\leftarrow}{\partial} \over \partial p}
{\stackrel{\rightarrow}{\partial} \over \partial q})
}
B(p,q),
\eea
where
$
S=\det \pmatrix{1 & 1 & 1\cr q& \xi& \xi'\cr p&\eta&\eta'\cr}.
$

Of course, the non-locality of a star product is a consequence of the
ordinary quantum-mechanical non-locality.

In the phase space parameterized by coordinates
$\gamma$ we have an analogue of algebra (\ref{alg}) with
the {\em Moyal bracket}
\be\label{algmoy}
[A,\,B]_{\rm MB} =
A\star B-B\star A=
i\hbar \{A,B\}+O(\hbar^2),
\ee

So far the star product is defined only by  the
relations (\ref{alg}). Unfortunately, all these
results are formal in the sense that they do not offer a
receipt for the procedure of the star-product construction.
The basic problem in attempt to generalize the exponentiation idea
(\ref{hstar})
to a non-constant Poisson structure is that
$\stackrel{\leftarrow}{\partial},\,\stackrel{\rightarrow}{\partial}$
no longer commute with the $\omega$. Nevertheless, recall that for
$\hat{A},\hat{B}$ in some Lie algebra and for
$\exp (\hat{A}) \exp (\hat{B})= \exp (\hat{C})$ the
Campbell-Baker-Hausdorff formula allows to define
$\hat{C}$ as a formal series whose terms are elements in
the Lie algebra generated by $\hat{A},\hat{B}$.
The associativity of such a way defined star product is induced
from the associativity of the group multiplication.

It is essential that the star
product determines  the higher $O(\hbar^2)$ terms up to gauge
equivalence, which amounts to linear redefinitions of the functions
$$
A'(\gamma)=A(\gamma)+\hbar S_{1}(A)+\hbar^2 S_{2}(A) + \cdots =(S(\hbar)A)(\gamma),
$$
with $S_{i}$ being differential operators. Two star products
related to each other by $S$ so that $S(A\star
B)=S(A)\star^{'}S(B)$ for all $A,B$ may therefore be considered
equivalent. Imposing associativity will constrain this operation and 
determine higher terms.

The multiplication kernel in the integral form and the local
form of {\em star operator}
$U= U(\stackrel{\leftarrow}{\partial}_{\gamma},
\stackrel{\rightarrow}{\partial}_{\gamma}; \omega)$
can be found from the Stratonovich-Weyl kernel in principle.
At least in some special cases the kernel satisfies the Schr\"odinger
equation, where the role of time is played by the {\em noncommutativity
parameter} $\hbar$ and  the role of a hamiltonian is played by the
Poisson structure associated with the deformation \cite{Ber}.
In cases under interest the star operator $U$ has
an exponential form of a one-parameter group element, like for the
flat case, i.e.
$U=\exp (i\hbar\Delta (\stackrel{\leftarrow}{\partial}_{\gamma},
\stackrel{\rightarrow}{\partial}_{\gamma}; \omega))$.
It should be noted that both integral and differential versions
of a star product allow for the {\em quasiclassical
expansion} of the composition law for symbols in power series
of the noncommutativity parameter $\hbar$ of algebra
(\ref{alg}).

As we have already pointed out 
in this section the symbol $\leftrightarrow$ operator
correspondence is a mathematical quantization problem 
finding of spectrum of the operators. The technique of operator symbols
allows to reformulate this problem from operator language on a language 
of the functions defined on some (classical) phase space with specific
non-commutative product rule (star product). From this point of view
the star operator is nothing but a quantum object. This means
that we have to solve a quantum problem (i.e. find eigenvalues
and eigenfunctions on a manifold) for a particular operator in
order to construct star operator exactly leaving aside problems of 
convergence and of construction of the Hilbert space. 
This scheme is generalized to the quantization of any symplectic or Poisson
manifold and the problem of existence and classification up to
equivalences formal star product was solved by several authors
(see for example \cite{Stern}) who support the belief that the formal 
deformation encloses the essential information 
of the quantum system.
The main direction, which has resulted in a simple
geometrical construction based on the Weyl algebras bundle,
consists in an observation that each tangent space of a
symplectic manifold is a symplectic vector space, so it can be
quantized by the usual Moyal-Weyl product \cite{Fed}.  Any
tangent space $T_{x}M^n$ can be parameterized by {\em normal
coordinates} $y$ which are transferred to the base space $M^n$ by
the exponential map of the given connection $\Gamma$ at $x$.  It
was found an iteration method of constructing a flat connection
on the Weyl bundle.  Intuitively, the flat connection can be
thought of as a quantum correction to the usual affine connection
on the tangent bundle and one may consider quantization procedure
developed in \cite{Fed} as a way of constructing a {\em quantum
exponential map} which always exists \cite{wein}. Any star
product is gauge equivalent to 
\be\label{qem} 
(A\star B)= \left[
(\exp^{\star}_{x}A)(y)\star_{\hbar}(\exp^{\star}_{x}B)(y)
\right]|_{y=0},
\ee
where $\exp_{x}: T_{x}M^n \rightarrow M$ is the exponential map,
defined in a neighborhood of the origin, corresponding to the
connection $\nabla$ and the $\star_{\hbar}$ refers to
standard Moyal-Weyl star product on symplectic vector space $T_{x}M^n$.
Below, it will be presented several significant explicit examples that
demonstrate practical realization of Eq (\ref{qem}).

Star product formulae (\ref{hstar}) are
very unhandy for practical computations. But, since the differential
form involves an exponential of derivative operators, it may be evaluated
through translation of function arguments. This can be easily seen
from (\ref{hstar}), which might be rewritten in the following form
\be
(A\star B)(p,q)= A(p-{i\over 2}\hbar{\partial \over \partial q_1},
q+{i\over 2}\hbar{\partial \over \partial p_1})
B(p_1,q_1)|_{p_1=p, q_1=q}=
A(p_{\hbar},q_{\hbar})B(p_{\hbar},q_{\hbar})\times 1,
\ee
where quantities
$
p_{\hbar}=p-{i\over 2}\hbar\stackrel{\rightarrow}{\partial_q} \,
q_{\hbar}=q+{i\over 2}\hbar\stackrel{\rightarrow}{\partial_p}
$
were introduced. They can be considered as a right regular
representation of generating operators $\hat{P}, \hat{Q}$.

In the general case, we can rewrite expression (\ref{hstar}) in a
more symmetrical form
\bea
(A\star B)(\gamma) &= &1 \times {\rm e}^{i\hbar\hat{\Delta}
(\stackrel{\leftarrow}{\partial}_{\gamma},
\stackrel{\rightarrow}{\partial}_{\gamma})}
A(\gamma)
{\rm e}^{-i\hbar\hat{\Delta}
(\stackrel{\leftarrow}{\partial}_{\gamma},
\stackrel{\rightarrow}{\partial}_{\gamma})}
{\rm e}^{i\hbar\hat{\Delta}
(\stackrel{\leftarrow}{\partial}_{\gamma},
\stackrel{\rightarrow}{\partial}_{\gamma})}
B(\gamma)
{\rm e}^{-i\hbar\hat{\Delta}
(\stackrel{\leftarrow}{\partial}_{\gamma},
\stackrel{\rightarrow}{\partial}_{\gamma})}\times 1 =\nonumber \\
 &=& \stackrel{\rightarrow}{A}_{\hbar}(\gamma)
 \stackrel{\rightarrow}{B}_{\hbar}(\gamma)\times 1 =
 A(\stackrel{\rightarrow}{\gamma_{\hbar}})
 B(\stackrel{\rightarrow}{\gamma_{\hbar}})\times 1 = \label{rep} \\
 &=& 1\times\stackrel{\leftarrow}{A}_{\hbar}(\gamma)
 \stackrel{\leftarrow}{B}_{\hbar}(\gamma) =
  1\times A(\stackrel{\leftarrow}{\gamma_{\hbar}})
 B(\stackrel{\leftarrow}{\gamma_{\hbar}}). \nonumber
\eea

Operators
$
\stackrel{\leftarrow}{\gamma}_{\hbar}, \,
\stackrel{\rightarrow}{\gamma}_{\hbar}
$ are left and right regular representation operators
$\hat{\gamma}$ with commutation relations (\ref{alg}).
The recipe
\be\label{trans}
A \rightarrow A_{\hbar}=U^{-1}AU
\ee
in the geometric quantization aspect is just a {\em prequantization procedure}
for the phase-space $X=(\gamma,\omega)$. The relation between symbols
and operators $A_{\hbar}$ has a very simple form
\be\label{bridge}
A = \stackrel{\rightarrow}{A_{\hbar}}\times 1 =
1\times\stackrel{\leftarrow}{A_{\hbar}}
\ee
This rule just means that all derivatives that act on nothing
must be omitted.

For our purposes, the most important formula, which takes place
in the framework of the symbol calculus due to the special
properties of the Stratonowich-Weyl kernel, is the trace definition
\be\label{symtr}
{\rm Tr}\hat{A} = \int_{X}d\mu(\gamma)\,A(\gamma),
\ee
where $X=(\gamma, \omega)$ is the phase space with invariant measure
$d\mu$ and $A(\gamma)$ is some symbol of operator $\hat{A}$.
It shold be noted that the definition (\ref{symtr})
is correct for any allowed choice of the Stratonowich-Weyl kernel
and for any ordering prescription in the enveloping algebra (\ref{alg}).
According to Eq (\ref{symtr}) finding of trace of the
operator $\hat{A}$ is reduced to constructing the corresponding
symbol $A(\gamma)$.
For example
$$
\sigma(\exp \hat{A})=\sum {1 \over n!} A\star A\star\ldots\star A
$$
is a symbol of an evolution operator of a some quantum-mechanical
problem, and to find the trace of corresponding operator one have
to perform the integration of the symbol.  However, for the
operators having intricate structure, obtaining its symbol via
explicit evaluation of Wigner's mapping appear intractable.  The
special representation $\gamma_{\hbar}$ of operators
$\hat{\gamma}$ appears to be more suitable.

\subsection{Examples of Star Product}
The last and the principal question is how to construct and compute
a concrete star operator $U$ on a special phase space.
Let us cite several well-known solutions of this problem which
demonstrate practical realization of formula (\ref{qem}) on special
symplectic structures. Let us consider some well-known examples of star product explicitly.
As we will see these examples allow to clarify a principial
possibility introducing the star product construction in superspace.

The first example we take from ref. \cite{BT}.
In those paper, it was shown that the deformation quantization yields to a
noncommutative algebra of functions (\ref{algmoy}) for each
Poisson-Lie structure on the arbitrary symplectic manifold $M$,
both in the nondegenerate and degenerate cases in the presence of
 the second-class constrains. In the nondegenerate case, we take
the Darboux coordinates in the initial phase space as a local
model, whereas in the degenerate case the same role of special
coordinates is played by physical variables on the constraint
surface. In this case one must consider Dirac's brackets as
a classical limit of brackets (\ref{algmoy}).
For the wide range of commutation relations (\ref{alg})
the formal scheme has been found and consists in replacement
partial derivatives by covariant ones $\partial_{\gamma}\rightarrow\nabla_{\gamma}$
in the Groenewold's noncommutative star product (\ref{hstar}),
which is manifestly associative.
This scheme allows one to avoid direct reducing the dynamic on the
curved shell, because such a reduction usually breaks explicit
covariance and space-time locality.

In the second example, motivated by the flat case $T^{*}R^{n}$,
authors \cite{Bord} constructed
homogeneous star products of Weyl and standard ordered type on
every cotangent bundle $T^{*}Q$ by means  of the Fedosov procedure
using a symplectic torsion-free connection. Their result presents
a surprisingly natural analogue of the operator $U$, which takes
the form $U=\exp ({\hbar \over 2i}\Delta)$, where the second order
differential operator $\Delta$ is equal to
$$
(\Delta)={\partial^2 \over \partial q^{i}\partial p_{i}} +
\Gamma^{j}_{ji}(q){\partial \over \partial p_{i}} +
p_{k}\Gamma^{k}_{ij}(q){\partial^2 \over \partial p_{i}\partial p_{j}}+
\alpha_{i}(q){\partial \over \partial p_{i}}.
$$
Here $\Gamma^{i}_{jk}$ are the Christoffel symbols of
the connection $\nabla$ and $\alpha_{i}$ is a particular choice
of a one-form on $Q$ such that $-d\alpha$ is equal to the trace
of the curvature tensor. For the Levi-Chivita connection of a
Riemanian metric $\alpha=0$.

The last example (see ref. \cite{Kar} and references therein) gives
the solution of the problem of gauge invariance in the
classical-quantum correspondence. Let us consider an abelian gauge theory.
The most natural conjecture is to replace the
gauge dependent canonical momentum $\hat{P}'$ entering
definition (\ref{swks}) by the gauge invariant kinetic momentum
$\hat{P}= \hat{P}'+ A(\hat{Q})$. Obviously, the product rule
for gauge invariant Weyl symbols will be different from the usual
Moyal product.
The algebra of operators (\ref{alg}) is
\be\label{algqed}
[\hat{Q}^{i},\hat{Q}^{j}]=0,\;
[\hat{Q}^{i},\hat{P}_{j}]=i\hbar\delta^{i}_{j},\;
[\hat{P}_{i},\hat{P}_{j}]=i\hbar F_{ij}(\hat{Q})
\ee
The "magnetic" star product $\star_{\rm F}$ corresponding
to commutation relations (\ref{algqed}) can be calculated by
the formula
\be\label{starqed}
(A\star_{\rm F}B)(q,p)=
A(q,p)U_{\rm F}B(q,p),
\ee
where star operator is 
\be\label{mstar}
U_{\rm F}=\exp ({i \over \hbar}\phi(q,i\hbar\stackrel{\leftarrow}{\partial}_{p},
i\hbar\stackrel{\rightarrow}{\partial}_{p})+
{i\hbar \over 2} (
\stackrel{\leftarrow}{\partial}_{q}\stackrel{\rightarrow}{\partial}_{p}-
\stackrel{\leftarrow}{\partial}_{p}\stackrel{\rightarrow}{\partial}_{q}
)),
\ee
and the phase $\phi$ defined as
$$
\phi (q,u_2,u_1)=\int_{0}^{1}ds\,\int_{0}^{s}dt\;
u_2 F(q+(s-{1\over 2})u_1+(t-{1\over 2})u_2)u_1.
$$
The first two terms of the magnetic star product expansion are
$$
A\star_{\rm F}B= AB-{i\hbar \over 2}\{A,B\}_{\rm F}+O(\hbar^2),
$$
where $\{\,,\,\}_{\rm F}$ is the Poisson bracket corresponding
to the symplectic form
$\omega_{\rm F}=\omega_{0}+{1 \over 2}F_{ij}(q)dq^{i}\wedge dq^{j}$,
i.e.
$$
\{A,B\}_{\rm F}=
\partial_{p}A\partial_{q}B-
\partial_{q}A\partial_{p}B+
F_{ij}\partial_{p_{i}}A\partial_{p_{i}}B.
$$
The main object we are interested in is the star operator $U_{\rm F}$ (\ref{mstar}).
Using transformation (\ref{trans}) of appropriate symbols one can
easily obtain all objects that are needed for algebra (\ref{algqed}).
For a kind of $QP$ ordering
prescription the star operator has the form
$U={\rm e}^{i\partial_{p}\cdot\nabla}$
(see for example ref. \cite{we}) and leads to the right regular representation of the
kinetic momenta \be\label{qpqed} \nabla^{\hbar}_{\mu}= ip_{\mu}+
i\int^{1}_{0}d\tau\,\tau F_{\nu\mu}(x+i\tau\partial_{p})
\partial^{\nu}_{p},
\ee
which is a the normal coordinate expansion over
$\partial_{p}\in T^{*}M$. It should be noted that we have never used
gauge fixing, but the second term in the right side of (\ref{qpqed})
is nothing but the vector potential in the Fock-Schwinger gauge
"$y^{\mu}A_{\mu}(x+y)=0$".
This particular example shows us that we can use covariant quantities
from the very beginning, which leads us to a modified star product.
Summarizing the above-stated facts,
it can be noticed that the well-known ad-hoc quantization rules
on cotangent bundles are obtained by the deformation quantization
in a very systematic way.

\subsection{Application of Symbol Technique to Superfield Models}
Let us discuss how the above technique  can be immediately applied to
supersymmetric field theories formulated in N=1 superspace. 
Considered in the previous section examples of star product operator 
for nontrivial symplectic manifolds with and general construction of 
quantum exponential map (\ref{qem}) allow us to give a good definition 
of Moyal-Weyl deformations of phase superspace.

First we observe  that $X=(M^n, T^{*}M, \omega)$ in (\ref{symtr}) is a
symplectic manifold. For each given point in the base manifold
$x\in M^n$ we have all possible tangents $y=\partial_{p}$ lying in the
tangent space $TM$ which is a {\bf flat space}. Therefore,
using the {\em horizontal lift} of the derivative operator in the
tangent bundle allows one to transfer immediately the flat definitions
into covariant and gauge invariant ones even on a superspace,
because the SUSY algebra is an example of a linear graded Poisson structure.
The exact determination of this algebra that corresponds to the quantization 
of systems with both, bosonic and fermionic degrees of freedom was given 
in ref. \cite{mar}.
In the previous section we saw the general structure of star operator. Common 
sense suggests us that the structure of a star operator on a superspace should 
match the considered examples with minimal changes related to a specificity of 
the supersymmetry algebra. 
In fact, a star operator $U$ on flat phase superspace was first
introduced in ref. \cite{jain}. Direct its calculation on superspace
can be found in ref. \cite{we}.

Since practical purpose consists in finding the trace
${\rm Tr}(\rm e^{H(\hat{\gamma})})$, we can summarize the steps needed:
\begin{itemize}

\item Construct {\em star operator} $U$ using the commutation
relation of basis operators $\hat{\gamma}$ and calculate
exactly or approximately a regular representation
$\gamma_{\hbar}$.

\item Replace all operators $\hat{\gamma}$ in ${\rm e^{H(\hat{\gamma})}}$
by their special representation $\gamma_{\hbar}$ and fulfil ordering
(disentangling $\partial_{p}$ to make them acting on nothing)
in ${\rm e^{H(\hat{\gamma})}}$ in order to find the symbol of
$\sigma({\rm e^{H(\hat{\gamma})}})$.

\item Implement the integration $\sigma({\rm e^{H(\hat{\gamma})}})$ over
phase space with the measure $d\mu(\gamma)$.

\end{itemize}
Let us note that for our purposes (finding the trace) we need to know
neither ordering prescription nor the Stratonowich-Weyl kernel
but algebra (\ref{alg}) only. Therefore, we always can
choose the most preferable ordering prescription.
All obtained formulae depend on the symplectic structure
determined by the algebra and therefore, all results are
{\bf gauge independent}.

We consider now the supersymmetric gauge
theories in N=1 superspace. As well known the basic objects of all such
theories are chiral superfields of matter, superfield strengths
$W_{\alpha}$ and its conjugate and supercovariant derivatives
satisfying the algebra
\bea
&  \{\nabla_{\alpha},\nabla_{\dot{\alpha}}\}=i\nabla_{\alpha\dot{\alpha}},
\; \{\bar{\nabla}_{\dot{\alpha}}, \nabla_{\beta\dot{\beta}} \}=
\epsilon_{\dot{\alpha}\dot{\beta}}W_{\beta},\;&\nonumber\\
 & [i\nabla_{\alpha\dot{\alpha}},i\nabla_{\beta\dot{\beta}}]=
i\epsilon_{\alpha\beta}f_{\dot{\alpha}\dot{\beta}}+
i\epsilon_{\dot{\alpha}\dot{\beta}}f_{\alpha\beta}, &\label{lie}\\
\eea
which along with relations
\be\label{pois}
\{\nabla_{\alpha},\theta^{\beta}\}=\delta^{\beta}_{\alpha},\,
\{\bar{\nabla}_{\dot{\alpha}},\bar{\theta}^{\dot{\beta}}\}=\delta^{\beta}_{\alpha},\,
[\nabla_{\alpha\dot{\alpha}},x^{\beta\dot{\beta}}]=
\delta^{\beta}_{\alpha}\delta^{\dot{\beta}}_{\dot{\alpha}}
\ee
provide the obvious Poisson-Lie superalgebra inherited
connection with flat torsion.

Let us apply  previously described receipts to find
right regular representation of superspace derivatives.
Firstly, we introduce the following notations for symbols of
flat derivative operators
\be
\sigma({\partial \over \partial \theta} )=\psi, \,
\sigma({\partial \over \partial \bar{\theta}} )=\bar{\psi}, \,
\sigma(-i{\partial \over \partial x} )= p.
\ee
Then the covariant superspace derivatives will have the
following symbols
\be
\sigma (\nabla_{\alpha})=\left( \psi- {1 \over 2}\theta p + {\cal A} \right)_{\alpha},\,
\sigma (\bar{\nabla}_{\dot{\alpha}})=\left( \bar{\psi}- {1 \over 2}p\bar{\theta} +\bar{\cal A} \right)_{\dot{\alpha}},\,
\sigma (\nabla_{\alpha\dot{\alpha}})=
\left(ip + {\cal A} \right)_{\alpha\dot{\alpha}}
\ee
where ${\cal A}$ stands for the connection.

As it was mentioned above the algebra of operators
(\ref{alg}) leads to a star product of their symbols.
We transform symbols to the right regular representation
in order to use the prescription (\ref{rep}) for the star product
calculation. First of all the star operator $U$ has to be
constructed. The way to construct star operator $U$ is to
direct transfer the basic definitions (\ref{swks}, \ref{rule}) and
(\ref{hstar}) in a phase superspace. It can be obtained also by analogy
with examples considered because the supersymmetry algebra (\ref{lie})
is a particular case of (\ref{alg}).

For further use it is convenient to introduce two
non-symmetric chiral forms for the star operator $U$.  These
forms are associated with suitable heat kernels $K^{-}$ and
$K^{+}$ which will be discussed in more details in the Section
\ref{kernels}. 
Let's define the following star operators
\bea
{\rm for}\; K^{-}:&U={\rm e}^{-\bar{\partial}_{\bar{\psi}}\bar{\nabla}} {\rm e}^{{1 \over
2}(\partial_{\psi} p \bar{\theta}-\theta p
\bar{\partial}_{\bar{\psi}})} {\rm e}^{-\partial_{\psi}\nabla}
{\rm e}^{-i\nabla\partial_{p}},&\nonumber\\
{\rm for}\; K^{+}:&U= {\rm e}^{-\partial_{\psi}\nabla}
{\rm e}^{{1 \over 2}(\partial_{\psi} p \bar{\theta}-\theta p
\bar{\partial}_{\bar{\psi}})}
{\rm e}^{-\bar{\partial}_{\bar{\psi}}\bar{\nabla}}
{\rm e}^{-i\nabla\partial_{p}}.\label{u}
\eea
Such form of star operators can be obtained by general method described above.
These forms of U correspond to some special ordering
prescription for operator product.  The presence of exponential
${\rm e}^{{1 \over 2}(\partial_{\psi} p \bar{\theta}-\theta p
\bar{\partial}_{\bar{\psi}})}$ in (\ref{u}) related to the
specificity of supersymmetry algebra and sevres to guarantee covariance.
The different expressions for U (\ref{u}) are stipulated by different choice of the phase
superspace coordinates. Using operators (\ref{u})  and
transformation rule (\ref{trans}), one can find the right regular
representation (\ref{lie}). The operators $\gamma_{\hbar}$ are
written as power series in normal coordinate system of vector bundles,
where role of coordinates in the tangent space $y$ are played by
right derivatives $\partial$ with coefficients which are superspace derivatives
of strength fields.
For calculation of the heat kernel $K^{-}$ we find
\bea\label{hder}
\bar{\nabla}^{\hbar}_{\dot{\alpha}}&=&\bar{\psi}_{\dot{\alpha}},\nonumber
\\
\nabla^{\hbar}_{\alpha}&=&\psi_{\alpha}-p_{\alpha\dot{\alpha}}\bar{\partial}^{\dot{\alpha}}
+ {i\over 2}\bar{\partial}^{\dot{\alpha}}(\partial^{\beta}_{\dot{\alpha}}f_{\beta\alpha}+
\partial^{\dot{\beta}}_{\alpha}f_{\dot{\beta}\dot{\alpha}})+
i\bar{\partial}^2 W_{\alpha} + i\dot{\partial}^{\dot{\alpha}}
\partial_{\alpha}\bar{W}_{\dot{\alpha}} +\\
& & +i\bar{\partial}^2 \partial^{\beta}f_{\alpha\beta} +
\bar{\partial}^2 \partial_{\alpha}D' + \bar{\partial}^2 \partial^2
\nabla^{\dot{\alpha}}_{\alpha}\bar{W}_{\dot{\alpha}} +
\ldots,\nonumber\\ i\nabla^{\hbar}_{\alpha\dot{\alpha}} &=&
\{\nabla^{\hbar}_{\alpha}, \nabla^{\hbar}_{\dot{\alpha}}\}, \nonumber\\
\eea
where derivatives mean
$
\bar{\partial}^{\dot{\alpha}} ={\partial \over \partial\bar{\psi}_{\dot{\alpha}}},\,
\partial^{\alpha}={\partial \over \partial\psi_{\alpha}},\,
\partial^{\alpha\dot{\alpha}}={\partial \over \partial p_{\alpha\dot{\alpha}}}
$
and the dots stand for the number of apparent higher
derivative terms. For images of the material and gauge strength
fields transformed by quantum exponential map (\ref{trans}) we
keep the original notations
\bea
\Phi^{\hbar}&=&\Phi+\partial^{\alpha}\Psi_{\alpha}-\partial^2 F
+\ldots,\label{hfield} \\
W^{\hbar}_{\alpha}&=&
W_{\alpha}+\partial^{\beta}f_{\beta\alpha}- i\partial_{\alpha}D'-
i\partial^2(\nabla^{\dot{\alpha}}_{\alpha}\bar{W}_{\dot{\alpha}})
+\ldots, \nonumber
\eea
where superfields defined as $\Psi=\nabla\Phi, F=\nabla^2\Phi$ and
the definition of component fields is given in Appendix C.

The next stage of the trace exponential operator calculation is to rewrite all
operators in (\ref{kinet}) in the regular representation form
(\ref{hder},\ref{hfield}). The resulting $H_{\hbar}$
is used in expression (\ref{zeta}). After implementation
of the ordering procedure we have to perform an integration
with the measure
\be\label{measure}
d\mu={1\over (2\pi)^4}d^4 p\, d^2\psi\, d^2\bar{\psi}\, d^8z.
\ee
The next sections demonstrates the outlined program
in practice. Some of applying tricks allow us to calculate string $H\star H\star H \ldots$ 
star product nonperturbatively.

\section{The Heat Kernels and Effective action}\label{kernels}

\subsection{Splitting of the contributions}

We begin now an application  of general symbol technique to
evaluating the effective action in the hypermultiplet model couple to
external N=2 vector multiplet.

To find the functional trace (\ref{init}) for
model (\ref{act}) we will use $\zeta$-representation
(\ref{zeta}) with kinetic operator (\ref{kinet}).
Accordingly to the program described in the previous
section, the trace calculation consists in
replacement of the differential operators and superfields by
their regular representation (\ref{hder}), (\ref{hfield})
and integration over measure (\ref{measure}).
The details of this procedure are presented in the previous
section.

Operator (\ref{kinet}) includes covariant superspace
derivatives of the fields $W, \bar{W}$ and $\Phi, \bar{\Phi}$.
The expansion in powers of $\nabla\Phi$ determines
the auxiliary field potential. We will not touch this problem
in the present paper since it requires a special and independent
investigation. Background (\ref{backgr}) implies that all
derivatives of $\Phi$ vanish, so that we are left with
(\ref{kinet}) with constant fields $\Phi,\bar{\Phi}$, which play
the role of the "mass" parameter. In this approximation the heat kernel
$K({T/\mu^2})$ in (\ref{zeta})
can be placed in a separate exponential like it was done in
ref. \cite{ag} since diagonal and non-diagonal parts of
the matrix (\ref{kinet}) become commuting
$$
{\rm e}^{T\hat{H}}=\sum_{n=0}^{\infty}{T^{2n}\over (2n)!}
\pmatrix{
\Phi\bar{\Phi}\bar{\nabla}^2\nabla^2& 0\cr
0& \bar{\Phi}\Phi\nabla^2\bar{\nabla}^2\cr
}^n \exp
\pmatrix{
\bar{\nabla}^2\nabla^2 & 0 \cr
0&\nabla^2\bar{\nabla}^2\cr
}.
$$
Performing the two dimensional matrix trace, we obtain
$\zeta_{\hat{H}} = \zeta_{\hat{H}}^{+}+\zeta_{\hat{H}}^{-}$
with
\be\label{zetta}
\zeta_{\hat{H}}^{-}(s)=
\int d^8z\, d^8p\;
\int_{0}^{\infty}
{dT \over \Gamma (s)}
T^{s-1}\sum_{n=0}^{\infty} {T^{2n} \over (2n)!}
({\Phi\bar{\Phi} \over \mu^2})^n
{d^{n}\over dT^{n}}
[\exp ({{T \over \mu^2}\bar{\nabla}^2 \nabla^2})]_{\pc}
\ee
and
$
\zeta_{\hat{H}}^{+}=
\zeta_{\hat{H}}^{-}({\bar{\nabla}^2  \leftrightarrow  \nabla^2}).
$

The operators $\bar{\nabla}^2 \nabla^2$ and $\nabla^2\bar{\nabla}^2$
are equivalent to chiral and antichiral
D'Alambertians since they are
acting on subspaces of chiral and antichiral quantum superfields
in (\ref{kinet}) and therefore one can use in (\ref{zetta})
the following identities (see f.e. \cite{ggrs})
\bea
& & \nabla^2\bar{\nabla}^2 =
\Box_{+}=\Box-i\bar{W}^{\dot{\alpha}}\bar{\nabla}_{\dot{\alpha}} -
{i \over 2}(\bar{\nabla}\bar{W}),\nonumber \\
& & \bar{\nabla}^2\nabla^2 =\Box_{-}=
\Box-iW^{\alpha}\nabla_{\alpha}-{i \over 2}(\nabla W)
\eea
For a general background the separation of the Hilbert
space of the superfields on invariant subspaces for the
supersymmetry algebra representations is a highly non-trivial problem.
This problem can be solved by the observation that external field
action $S_{0}$ (\ref{act}) in N=1 form includes kinetic terms for
the $W$ and $\Phi$ fields, which involve the integration over
chiral and whole superspace measure. This observation suggests an
idea to present the quantum corrections in the same form.  So, in
order to separate out the contributions which renormalize each
kinetic term separately in the action $S_{0}$ we present
(\ref{zetta}) as a sum, which correctly determines corresponding
renormalizations. Expression (\ref{zetta}) is naturally rewritten
as a sum of two terms (as we will see later, this definition
gives the correct coefficients in leading terms)
\be\label{sum}
\zeta_{\hat{H}}^{\pm}=\zeta_{WW}^{\pm}+\zeta_{\Phi\bar{\Phi}}^{\pm}.
\ee
The part $\zeta_{WW}^{\pm}$ will include the renormalization of
the kinetic term $W^2$ in the classical action (\ref{act}) and
the part $\zeta_{\Phi\bar{\Phi}}^{\pm}$ includes the renormalization
of the kinetic term $\Phi\bar{\Phi}$. So, we have
\be\label{phi}
\zeta_{\Phi\bar{\Phi}}^{-}(s)=
\int d^8z\, d^8p\;
\int_{0}^{\infty} {dT \over \Gamma (s)}
T^{s-1}\sum_{n=1}^{\infty} {T^{2n} \over (2n)!}
({\Phi\bar{\Phi} \over \mu^2})^n
{d^{n-1}\over dT^{n-1}}
[\exp ({{T \over \mu^2}\Box_{-}}) \bar{\nabla}^2 \nabla^2]_{\hbar},
\ee
\be\label{ww}
\zeta_{WW}^{-}(s)=
\int d^6z\, d^8p\;
\int_{0}^{\infty} {dT \over \Gamma (s)}
T^{s-1}\sum_{n=0}^{\infty} {T^{2n} \over (2n)!}
({\Phi\bar{\Phi} \over \mu^2})^n
{d^{n}\over dT^{n}}
[\exp ({{T \over \mu^2}\Box_{-}}) \bar{\nabla}^2]_{\hbar},
\ee
with
$
\zeta^{+}=\zeta^{-}
(\nabla^2 \leftrightarrow \bar{\nabla}^2,
\Box_{+} \leftrightarrow \Box_{-}).
$

At the next step we evaluate the heat kernels (\ref{zeta})
for (\ref{ww}) and (\ref{phi}). Detailed analysis is given
in the Appendix A. The results have the form
\bea\label{kp}
 K_{\Phi\bar{\Phi}}^{-}(T)&=&
 K_{\rm Sch}(T)
\times
\nonumber \\
& & \times \{
1 + W^2 \bar{W}^2 {T^3 \over 3} \left({\sin (TG/2) \over TG/2}\right)^2
(
1 - {3 \over 4}\left({\sin (TG/2) \over TG/2}\right)^2 \times
 \\
& & \times(\lambda_{2}T\coth \lambda_{2}T + \lambda_{1}T\cot \lambda_{1} T)
)
\},\nonumber \\
K_{WW}^{-}(T) &= &
K_{\rm Sch}(T)
T^2 W^2
\left(
{\sin TG/2 \over TG/2}
\right)^2,\label{kw}
\eea
where $G=\lambda_{1}+i\lambda_{2}$ and $\lambda_{1,2}$ are the
electric and magnetic Maxwell superfields
(eigenvalues
$F_{\alpha\dot{\alpha}\beta\dot{\beta}}=
\epsilon_{\alpha\beta}f_{\dot{\alpha}\dot{\beta}}+
\epsilon_{\dot{\alpha}\dot{\beta}}f_{\alpha\beta}
$) in a special coordinate
basis related to the invariants
$(\lambda_{1}\pm i\lambda_{2})=-{1 \over 2}(F^2\pm F^{*}F)$.
The other heat kernels might be obtained via a simple substitution
$
K^{+} = K^{-}(G \leftrightarrow \bar{G}, \,
W \leftrightarrow \bar{W}).
$
The notation
\be
K_{\rm Sch} ={i \over (4\pi T)^2}
{\lambda_{1}T \lambda_{2}T \over \sinh T\lambda_{2} \sin T\lambda_{1}}
\ee
is used for known Schwinger kernel at coinciding points.

\subsection{Asymptotic expansion of Effective Action}

Let us consider the proper time expansion (\ref{deck}),
(\ref{decz}) for the obtained kernels (\ref{kw}), (\ref{kp}) and
investigate various terms of effective action.

\subsubsection{Divergent contributions}
First terms in the decomposition (\ref{deck}) for all heat kernels
are divergent, so we will treat them one by one. As it is shown in
the Appendix B, the first coefficient for
$K_{\Phi\bar{\Phi}}^{-}$ corresponding to $f=0$ is
$$
\zeta_{0} (s) =
{\Phi\bar{\Phi} \over 2(4\pi)^2}
\left({\Phi\bar{\Phi} \over 4\mu^2}\right)^{-s}
\left({\sqrt{\pi}\over 2} {\Gamma (1-s) \over \Gamma (3/2-s)}\right).
$$
Implementation of $\zeta'(0)$ leads to the known K\"ahlerian potential
in EA (see f.e. \cite{bky})
\be\label{pd}
\left(\Gamma_{\Phi\bar{\Phi}}\right)_{\rm div} = \int d^8z\,
{\Phi\bar{\Phi} \over 32 \pi^2}
\left(2-\ln {\Phi\bar{\Phi} \over \mu^2} \right).
\ee
which gives rise to the holomorphic Seiberg type
effective potential
$\bar{\Phi}{\cal F}'(\Phi)$.
For the heat
kernel $K_{WW}^{-}$ the first term corresponding to $f=2$ is \be
\zeta_2 (s) = {1 \over 2}
\left({\Phi\bar{\Phi}\over 4\mu^2 }\right)^{-s},
\ee
which gives the standard divergent and the holomorphic scale dependent
contribution
\be\label{wd}
\left(\Gamma_{WW}\right)_{\rm div} = -{1 \over 2 (4\pi)^2}\int d^6z\,
W^2\ln {\Phi \over \mu}.
\ee

It is easy to see that divergent contributions (\ref{pd}) and
(\ref{wd}) may be combined together to make the holomorphic part of the N=2
EA (\ref{n2})
\be\label{sp}
{\cal F}({\cal W})= -{1 \over (4\pi)^2}{\cal W}^2
\ln {{\cal W}^2 \over \mu^2}.
\ee
Equation (\ref{sp}) is well known Seiberg type low-energy
effective potential for the model under consideration.

\subsubsection{Finite contributions}

Other terms in the heat kernel decomposition
(\ref{deck}), (\ref{decz}) give finite contributions and
correspond to the inverse mass decomposition form
\be\label{wf}
\left(\Gamma_{WW}\right)_{\rm fin} =
{1 \over 2 (4\pi)^2}
\sum_{f} {\Gamma (f-2) \over (\Phi\bar{\Phi})^{f-2}}a_{f}
\ee
for the kernel $K_{WW}^{-}$ and
\be\label{pf}
\left(\Gamma_{\Phi\bar{\Phi}}\right)_{\rm fin} =
- {1 \over 4 (4\pi)^2}
\sum_{f} {\Gamma (f-1) \over (\Phi\bar{\Phi})^{f-1}}a_{f}
\ee
for the kernel $K_{\Phi\bar{\Phi}}^{-}$, where the coefficients
$a_f$ are obtained from decomposition (\ref{deck}) for
the corresponding kernels.

>From (\ref{wf}) and (\ref{kw}) we can obtain
\be
\left(\Gamma_{WW}^{-}\right)_{\rm fin}=
{1 \over 2 (4\pi)^2}\int d^6z\,\int_{0}^{\infty} dt\,t{\rm e}^{-t}
{W^2 y^2 \over (\Phi\bar{\Phi})^2}\zeta(t\bar{\Psi},t\Psi),
\ee
where the function $\zeta(x,y)$ and quantities $t\bar{\Psi},
t\Psi$, which transform as scalars with respect to N=1
superconformal group, introduced in \cite{bk} were used
$$
\zeta(x,y)= { x^2 (\cosh y -1)-y^2
(\cosh x -1) \over x^2 y^2 (\cosh x -\cosh y) }.
$$

Taking into account another kernel $K^{+}_{\bar{W}\bar{W}}$,
we find the non-holomorphic form of the whole contribution
\be\label{wwfin}
\left(\Gamma_{W\bar{W}}\right)_{\rm fin}=
{1 \over (4\pi)^2}\int d^8z\,\int_{0}^{\infty} dt\,t{\rm e}^{-t}
{W^2 \bar{W}^2 \over (\Phi\bar{\Phi})^2}\zeta(t\bar{\Psi},
t\Psi).
\ee

Equations (\ref{pf}) and (\ref{kp}) lead to the following result. 
\bea\label{gp}
\left(\Gamma_{\Phi\bar{\Phi}}^{-}\right)_{\rm fin}&= &
{1 \over 4 (4\pi)^2}\int d^8z\,
\int_{0}^{\infty} {dt \over t^2}{\rm e}^{-t}
\Phi\bar{\Phi}\xi(t\bar{\Psi},t\Psi) - \nonumber\\
&-&
{1 \over 12 (4\pi)^2}\int d^8z\,\int_{0}^{\infty} dt\,t{\rm e}^{-t}
{W^2\bar{W}^2 \over (\Phi\bar{\Phi})^2}
\lambda(t\bar{\Psi}, t\Psi)\tau(t\bar{\Psi},t\Psi),
\eea
where
$$
\lambda(x,y)={ x^2-y^2 \over \cosh x-\cosh y }
{\cosh x-1 \over x^2},
$$
$$
\xi(x,y)=
{
(\cosh x -1 -x^2 /2)-(\cosh y -1 -y^2/2)
\over
\cosh x- \cosh y
},
$$
$$
\tau(x,y)=
{1- {3 \over 2}{\cosh y -1 \over y^2}
(
{y\sinh y - x \sinh x \over
\cosh y -\cosh x}
)}.
$$

Another chiral kernel can be obtained by the replacement
$
\left(\Gamma_{\Phi\bar{\Phi}}^{+}\right)_{\rm fin}=
\left(\Gamma_{\Phi\bar{\Phi}}^{-}\right)_{\rm fin}
(\Psi\leftrightarrow \bar{\Psi}).
$
According to the expression (\ref{sum}), the whole
set of finite contributions is obtained as a sum
\be\label{res}
\left(\Gamma\right)_{\rm fin}=
\left(\Gamma_{W\bar{W}}\right)_{\rm fin} +
\left(\Gamma_{\Phi\bar{\Phi}}\right)_{\rm fin}.
\ee

Eq (\ref{res}) is our final result.

As one can see the low-energy effective action (\ref{res})
contains the contributions of two different types. First, the
contributions of the (\ref{wwfin})-type. These terms are analogous
to ones given in \cite{bk} although they were obtained completely
another method. Such terms can be rewritten in manifest N=2
superconformal manner using the proper N=2 superconformal
invariant functional \cite{bk}. Another type of contributions has the
structure (\ref{gp}), it is quite new and never been investigated
before. The corresponding terms are manifestly N=1 superconformal
invariant however, in fact, they are invariant under N=2
superconformal transformations (see the transformations in
\cite{kt}).
Technique of building the manifest N=2 superconformal invariants
on the base of their N=1 projections was described in \cite{bk}.  The
crutial role is played by the superfields ${\boldmath \Psi}^{2}$ and
${\bar{\boldmath \Psi}}^{2}$ with simple transformation laws under N=2
superconformal group. We should also to introduce full N=2
superspace measure. As a result one can restore manifest N=2
superconformal structure of the (\ref{gp})-type contributions.
However it is necessary to point out that in process of such
restoration we have to use the derivatives of the N=1 superfields
$W_{\alpha}$ and ${\Phi}$ a final expression will certainly
contain the terms with higher derivatives of the N=2 strengths
${\cal W}$ and conjugate which are next to leading in compare with
terms kept in \cite{bk}. Thus the method used here allows in principle
to go beyond the results \cite{bk} and obtain the new N=2
superconformal invariant contributions in derivative expansion of
low-energy effective action.

\section{Conclusion}
Let us sum up the results of the paper. We have investigated a structure 
of induced effective action in hypermultiplet theory coupled to
	external N=2 vector multiplet realizing this theory in terms of N=1
superfields. Such an effective action is typical for N=1 chiral superfield
models. 

Within Schwinger proper-time method a calculation of the effective
action is reduced to mathematical problem of evaluating the
functional (super)trace of the exponent of superspace differential
operator associated with second variational derivative of initial classical
action. We have shown that this problem can be efficiently studied on the
base of technique of operator symbols adapted in this paper to SUSY
theories formulated in N=1 superspace. Use of such superfield operator
symbol technique allowed to develop a general procedure of
(supercovariant)derivative expansion of the effective action and calculate 
the leading contributions to low-energy effective action of the model
under consideration. 

We have found the low energy-effective action as a sum
of N=2 superconformal invariant terms (\ref{wwfin},\ref{gp}) constructed from
superconformal blocks introduced in \cite{bk} and the holomorphic
Seiberg-type terms violating N=2 superconformal symmetry. The result
obtained is most general up to now low-energy induced effective action for
hypermultiplet theory coupled to external N=2 vector multiplet.        

The approach developed in this paper can be applied to various problems 
associated with calculating effective action in superfield theories. As
the actual example we point out a general construction of derivative
expansion of effective action in quantum superconformal invariant theories 
including N=4 super Yang-Mills model.

\section{Acknowledgements}
The authors are extremely grateful to S.M. Kuzenko for discussing
the superfield structure of N=2 superconformal invariants and A.A.
 Tseytlin for valuable comments. The
work of I.L. Buchbinder was partially supported by RFBR grant
99-02-16617, DFG-RFBR grant 99-02-04022, INTAS grant 991-590,
GRACENAS grant 97-6.2-34 and NATO research grant PST.CLG 974965.
He thanks the Institute of Physics at the University of Sao Paulo
for hospitality and FAPESP for support.  This work was of
N.G. Pletnev and A.T. Banin supported by  RFBR 99-02-17211, RFBR
00-15-96691 grants and a grant of Sankt-Petersburg Center of
Fundamental Sciences.

\newpage
\section*{Appendixes}
\appendix

\renewcommand{\thesection}{Appendix \Alph{section}.}
\renewcommand{\theequation}{\Alph{section}.\arabic{equation}}

\setcounter{equation}{0}
\section{The Heat Kernel Calculation}

Let us consider the heat kernel corresponding to expression
(\ref{phi}).  In the constant field background (\ref{backgr}) it
contains an abelian field strength and its scalar superpartners
without higher derivatives.

The full integration measure (\ref{measure})on phase superspace in
(\ref{phi}) means that all nonzero contributions have to be
proportional at least to $\bar{\psi}^2$. The regular representation $\Box_{-}^{\hbar}$ in the exponent
will contain $\partial_{\dot{\alpha}}$, which will act on
$\bar{\nabla}^2_{\hbar}$. This action can only lower
degree of $\bar{\psi}$, which is a symbol of $\bar{\nabla}^2$.
So that, we should omit all $\partial_{\dot{\alpha}}$ in
$\Box_{-}^{\pc}$ and implement $d^2\bar{\psi}$ integration.
The Laplace-type operator $\Box_{-}^{\pc}$ becomes
more simple and we obtain
\be
K^{-}_{\Phi\bar{\Phi}}(T)=
\int {d^4p \over (2\pi)^4}d^2 \psi
[{\rm e}^{T(\stackrel{\circ}{\Box} +
\partial^{\alpha}\stackrel{\circ}{\nabla}_{\alpha\dot{\alpha}}\bar{W}^{\dot{\alpha}}
-\partial^2\bar{W}^2 +
i\psi_{\alpha}W^{\alpha}+
i\psi_{\alpha}\partial^{\beta}f^{\alpha}_{\beta}
)}\nabla^2]_{\pc},
\ee
with
$$
\stackrel{\circ}{\Box}_{\pc}={1 \over 2}
[\stackrel{\circ}{\nabla}^{\alpha\dot{\alpha}}
\stackrel{\circ}{\nabla}_{\alpha\dot{\alpha}}]_{\pc},\quad
\stackrel{\circ}{\nabla}_{\alpha\dot{\alpha}}^{\pc}=
ip_{\alpha\dot{\alpha}}+{1 \over 2}
(\partial^{\beta}_{\dot{\alpha}}f_{\beta\alpha} +
\partial^{\dot{\beta}}_{\alpha}f_{\dot{\beta}\dot{\alpha}}
).
$$
Using the known operator identity
$$
{\rm e}^{A+B}={\rm e}^A \exp (\int_{0}^{1} d\tau {\rm e}^{-\tau A}B{\rm e}^{\tau A})
$$
we disentangle of the Grassmanian momentum derivatives
$ \partial^{\alpha}$ to make them acting on nothing.
Then we implement a trivial integration over $d^2\psi$.
The result is
\be
K^{-}_{\Phi\bar{\Phi}}(T)=
\int {d^4 p \over (2\pi)^4}
{\rm e}^{T\stackrel{\circ}{\Box}}
\{
1 +{1 \over 3}T^3 \bar{W}^2\tilde{W}^2
-{1 \over 8}T^4 \bar{W}^2\tilde{W}^2 B^{\;\alpha}_{\beta}B^{\beta}_{\;\delta}
\nabla_{\alpha\dot{\alpha}}(T)\nabla^{\delta\dot{\alpha}}(T)
\},
\ee
here $\tilde{W}^{\alpha}=W^{\beta}\tilde{B}^{\;\alpha}_{\beta}$ and
\be
B^{\;\alpha}_{\beta}=
\left({{\rm e}^{-iTf}-1 \over -iTf}\right)^{\;\alpha}_{\beta}, \quad
\tilde{B}_{\beta}^{\;\alpha}=
\left({{\rm e}^{iTf}-1 \over iTf }
\right)^{\;\alpha}_{\beta},
\ee
$$
\nabla_{\alpha\dot{\alpha}}(T)=
{1 \over T}
\int_{0}^{T}d\tau\,({\rm e}^{-i\tau f})^{\;\delta}_{\alpha}
\nabla_{\delta\dot{\delta}}
({\rm e}^{-i\tau f})^{\;\dot{\delta}}_{\dot{\alpha}} =
\nabla_{\delta\dot{\delta}}
{\cal F}^{\delta\dot{\delta}}_{\alpha\dot{\alpha}}(T).
$$
The last step is the calculation of the standard Schwinger
heat kernel and moments $\langle \nabla_{A}\nabla_{B}\ldots\rangle $, i.e.
$$
K(T)_{AB\cdots}=\int {d^4 p \over (2\pi)^4}{\rm e}^{T\stackrel{\circ}{\Box}}
\nabla_{A}\nabla_{B}\cdots
$$
This problem can be solved with aid of an elegant
technique \cite{ag}. The method was used to determine the
moments $\langle \nabla_{A}\nabla_{B}\ldots\rangle $ in terms of the
Gaussian itself. To solve a differential equation
\be\label{equation}
{dK \over dT} = {1 \over 2}K_{\alpha\dot{\alpha}}^{\alpha\dot{\alpha}}
\ee
one can use the identity
$$
0=\int {d^4p \over (2\pi)^4}
\partial_{p}^{\alpha\dot{\alpha}}
({\rm e}^{T\stackrel{\circ}{\Box}}\stackrel{\circ}{\nabla}_{\delta\dot{\delta}})
$$
in order to write the expression for
$K_{\alpha\dot{\alpha}}^{\beta\dot{\beta}}$ in terms of $K(T)$ \be\label{kernel}
K_{\alpha\dot{\alpha}}^{\beta\dot{\beta}}=-K
{\cal F}^{-1 \,\beta\dot{\beta}}_{\alpha\dot{\alpha}}.
\ee
Now we obtain an explicit form ${\cal F}, {\cal F}^{-1}$ in terms of
Maxwell's invariants. Let us rewrite $F_{\mu\nu}$ in the special reference
frame
$$
F_{\mu\nu}=\pmatrix{
0 & \lambda_{1} & 0 & 0 \cr
-\lambda_{1} & 0 & 0 & 0 \cr
0 & 0 & 0 & -\lambda_{2} \cr
0 & 0 & \lambda_{2} & 0  \cr
},
$$
where $\lambda_{1,2}$ are related to the invariants
${\cal H}_{\pm}={1 \over 2}(F^2\pm i F^{*}F) = (\lambda_{1}\pm i\lambda_{2})^2$
Now, ${\cal F},\, {\cal F}^{-1}$ can be rewritten in the Pauli matrix
basis
$\sigma=(\sigma^{1})_{\alpha}^{\beta},\,
\bar{\sigma}=(\sigma^{1})_{\dot{\alpha}}^{\dot{\beta}}$
as
\bea
{\cal F}&=&{T \over 2}
(
({\sinh \alpha \over \alpha}+ {\sin \beta \over \beta}) +
\sigma ({\cosh \alpha -1 \over \alpha}+i{\cos \beta -1 \over \beta}) +
\nonumber \\
& & +\bar{\sigma}({\cosh \alpha -1 \over \alpha}-i{\cos \beta -1 \over \beta}) +
\sigma\bar{\sigma}({\sinh \alpha \over \alpha}-{\sin \beta \over \beta})
) \nonumber \\
{\cal F}^{-1}&=&{1 \over 4 T}
(
\alpha\coth {\alpha \over 2}+\beta\cot {\beta \over 2} +
\sigma (-\alpha +i\beta)+
\bar{\sigma}(-\alpha -i\beta)+
\\
& & +\sigma\bar{\sigma} (\alpha\coth {\alpha \over 2}-\beta \cot {\beta \over 2})
),\nonumber
\eea
where $\alpha=2T\lambda_{2}, \; \beta=2T\lambda_{1}$.

In this notation the solution of equation (\ref{equation}),
(\ref{kernel}) is
$$
K_{\rm Sch} ={C \over \sinh T\lambda_{2} \sin T\lambda_{1}}.
$$

By choosing the constant $C$ as ${i\lambda_{1}\lambda_{2}\over (4\pi)^2}$
which corresponds to standard boundary condition for $K(T)$ to be
reduced to the ordinary ${1 \over (4\pi T)^2}$, one finds the Schwinger result for
$K(T)$ and supersymmetric "corrections"
\be
K^{-}_{\Phi\bar{\Phi}}(T)=
K_{\rm Sch}(T)
\{
1 + {T^3 \over 3}\bar{W}^2\tilde{W}^2 +
{T^4 \over 8}\bar{W}^2\tilde{W}^2 B^{\,\alpha}_{\beta}
B^{\beta}_{\,\delta}({\cal F}^{-1})^{\delta\dot{\alpha}}_{\alpha\dot{\alpha}}
\}.
\ee
Direct computation of the traces over spinor indices leads to the full
heat kernel
\bea\label{kernphi}
 K^{-}_{\Phi\bar{\Phi}}(T)&=&{i \over (4\pi T)^2}
{\lambda_{1}T \over \sin \lambda_{1} T}
{\lambda_{2}T \over \sinh \lambda_{2} T}
\times
\nonumber \\
& & \times \{
1 + W^2 \bar{W}^2 {T^3 \over 3} \left({\sin (TG/2) \over TG/2}\right)^2
(
1 - {3 \over 4}\left({\sin (TG/2) \over TG/2}\right)^2 \times
 \\
& & \times(\lambda_{2}T\coth \lambda_{2}T + \lambda_{1}T\cot \lambda_{1} T)
)
\},\nonumber
\eea
where $G=\lambda_{1}+i\lambda_{2}$.

The calculations of the heat kernel for (\ref{ww}) is more
simple and lead to the expression
\bea\label{kernww}
K^{-}_{WW}(T) &=&\int {d^4p \over (2\pi)^4} d^2\psi d^2\bar{\psi}
[{\rm e}^{T\Box_{-}}\bar{\nabla}^2]_{\pc}=
\nonumber \\
&= &{i \over (4\pi T)^2}
{\lambda_{1}T \over \sin \lambda_{1} T}
{\lambda_{2} T \over \sinh \lambda_{2} T}
T^2 W^2
\left(
{\sin TG/2 \over TG/2}
\right)^2.
\eea

As one can easy to see in order to obtain the kernels $K^{+}$,
we should just replace $G\rightarrow\bar{G}$ in (\ref{kernphi}),
(\ref{kernww}).

\setcounter{equation}{0}

\section{Calculation of the $\zeta$-Functions}

\subsection*{Subleading terms of $\Phi\bar{\Phi}$}
Kernel (\ref{kernphi}) can be represented as a power series
in proper time $T$ (\ref{deck}), which starts with
term ${1 \over (4\pi T)^2}$.
Let us rewrite this term in the form
$$
{1 \over T^2}=\int_{0}^{\infty}dz\,z{\rm e}^{-zT},
$$
in order to remove derivatives ${d^{n-1} \over dT^{n-1}}$ in
(\ref{phi}). Consequently integrating by parts one can find
the appropriate decomposition (\ref{decz}) with the
coefficients $a_{n}$ which are defined by
$K^{-}_{\Phi\bar{\Phi}}(T)$ given in form (\ref{deck}).

We investigate the first term in (\ref{decz}) independently
because of its singularity. The simple manipulations lead
to the form
$$
\zeta_{(0)\; \Phi\bar{\Phi}}^{-}(s) =
{\Phi\bar{\Phi} \over 2(4\pi)^2}
{\Gamma (s+2) \over \Gamma (s)}
\left(\Phi\bar{\Phi} \over \mu^2 \right)^{-s}
\int^{\infty}_{0}
dz\, z^{s-1}
{_{3}F_{2}}(1,{s \over 2}+1, {s \over 2}+{3 \over 2};2,{3 \over 2};-z)
$$
which is the Mellin transform of the generalized
hypergeometric function ${_{3}F_{2}}$.

Using an integral representation
$$
{_pF_q}((a_p),(b_q);z)= {\Gamma (b_q) \over \Gamma (a_p) \Gamma (b_q-a_p)}
\int_{0}^{1}dt\,t^{a_p-1}(1-t)^{b_q-a_p-1}
{_{p-1}F_{q-1}}((a_{p-1}),(b_{q-1});tz)
$$
and
$$
\int_{0}^{\infty}dz\,z^{-p-1}{_2F_1}(a,b,c;-z)=
{\Gamma (a+p) \over \Gamma (a)}
{\Gamma (b+p) \over \Gamma (b)}
{\Gamma (c) \over \Gamma (c+p)}\Gamma (-p)
$$
we find
$$
\zeta_{(0)\; \Phi\bar{\Phi}}^{-} (s) =
{\Phi\bar{\Phi} \over 2(4\pi)^2}
\left({\Phi\bar{\Phi} \over 4\mu^2}\right)^{-s}
\left({\sqrt{\pi}\over 2} {\Gamma (1-s) \over \Gamma (3/2-s)}\right).
$$
This leads to the known K\"ahlerian potential \cite{bky}
\be
\left(\Gamma_{\Phi\bar{\Phi}}\right)_{\rm div} = \int d^8z\,
{\Phi\bar{\Phi} \over 32 \pi^2}
\left(2-\ln {\Phi\bar{\Phi} \over \mu^2} \right).
\ee
This result illustrates a correctness of the technique under
consideration.

The following terms of the decomposition in $(T/\mu^2)^f$ are
\bea
\zeta_{(f)\; \Phi\bar{\Phi}}^{-}(s)&= & {1 \over 2 (4\pi)^2}{\Phi\bar{\Phi} \over (\mu^2)^{-s}}
{\Gamma (s+2+f) \over \Gamma (s)}(\Phi\bar{\Phi})^{-s-f}\times
\nonumber\\
& & \times\int_{0}^{\infty}dz\,z^{s+f-1}
{_4F_3}(1,{s \over 2}+1,{s \over 2}+{3 \over 2},s+f+2,{3 \over 2},2,s+2; -z)
=
\nonumber \\
& & = {1 \over 2(4\pi)^2}{1 \over (\Phi\bar{\Phi})^{f-1}}
\left({\Phi\bar{\Phi} \over \mu^2} \right)^{-s}
{
\Gamma (1-f-s) \Gamma (f+s) \Gamma (2-s-2f) \over
\Gamma (s) \Gamma (1-f) \Gamma (3-2s-2f)
}.
\eea

Let us consider the terms with $f\neq 0$ in the $\zeta '(0)$.
To eliminate the zero at $s=0$ derivative ${d \over ds}$ must act only
on $1/{\Gamma(s)} \sim s$.  It leads to decomposition
\be
(\zeta_{(f)\;\Phi\bar{\Phi} }^{-})'(0)=-{1 \over 4 (4\pi)^2}{\Gamma (f-1) \over (\Phi\bar{\Phi})^{f-1}}
\ee
where the inverse power of $\Phi\bar{\Phi}$ plays the role of
the effective scale or mass.

\subsection*{Subleading terms for $W^2$}

Let us investigate another kernel (\ref{kernww}).
Once again we present $K^{-}_{WW}(T)$ as a series (\ref{deck}), i.e.
\be
K^{-}_{WW}(T)={1 \over (4\pi T)^2}\sum_{f=2}a_{f}T^{f}
\ee
and once again we obtain the Mellin transformation
\be
\zeta_{(f)\; WW}^{-}(s)=
\left({\Phi\bar{\Phi}}\over \mu^2 \right)^{-s}
(\Phi\bar{\Phi})^{2-f}
\int_{0}^{\infty}dz\, z^{s+f-3}
{\Gamma (s+f) \over \Gamma (s)}
{_3F_2}(s+f,{s\over 2},{s\over 2}+ {1\over 2}; s, {1 \over 2}; -z).
\ee
We obtain after the integration
\be
\zeta_{(f)\; WW}^{-}(s)=
\left(
{\Phi\bar{\Phi} \over 4\mu^2}
\right)^{-s}
\left(
{\Phi\bar{\Phi} \over 4}
\right)^{2-f}
{1 \over \Gamma (s)}
{\sqrt{\pi} \Gamma (s+f-2) \Gamma (-s-2(f-2)) \over
\Gamma (2-f) \Gamma (-s-f+5/2)
}.
\ee
If $f\neq 2$ then the only contribution to $\zeta '(0)$ is given
by $1/\Gamma (s)$ and the corresponding term is
\be
(\zeta_{(f)\; WW}^{-})'(0)={\Gamma (f-2) \over 2 (\Phi\bar{\Phi})^{f-2}}.
\ee
The case $f=2$ leads to the standard divergence
\be
\zeta_{(2)\; WW}^{-}(s) = {1 \over 2}\left({\Phi\bar{\Phi}\over 4\mu^2 }\right)^{-s}, \quad
(\zeta_{(2)\; WW}^{-})'(0)\sim -{1 \over 2}\ln {\Phi\bar{\Phi} \over 4 \mu^2}
\ee
and we obtain the known holomorphic scale dependent contribution.
The other terms in the decomposition (\ref{deck}) for
the heat kernel $K^{+}_{\bar{W}\bar{W}}$ certainly depend on $\bar{\Phi},
\nabla_{(\dot{\alpha}}\bar{W}_{\dot{\beta})}$ and we
have to recover the full superspace measure in (\ref{ww}),
i.e. $\int d^8z=\int d^6z\,\bar{\nabla}^2$ that leads to
expression (\ref{wwfin}).

\setcounter{equation}{0}
\section{Components}

We use the following component structure of the N=1 superfields
$W, \Phi$
$$
W=\lambda_{\alpha}+\theta^{\beta}f_{\beta\alpha}- i\theta_{\alpha}D'+
i\theta^2\partial_{\alpha\dot{\alpha}}\bar{\lambda}^{\dot{\alpha}}+
{i\over 2}\theta^{\beta}\bar{\theta}^{\dot{\beta}}\partial_{\beta\dot{\beta}}
\lambda_{\alpha} +{1 \over 4}\theta^{2}\bar{\theta}^{2}\Box\lambda_{\alpha}
-{i\over 2}\theta^2\bar{\theta}^{\dot{\beta}}\partial^{\beta}_{\dot{\beta}}
(f_{\alpha\beta}-iC_{\beta\alpha}D'),
$$
$$
\Phi=\phi+\theta^{\alpha}\psi_{\alpha}+\theta^{2}F +
{i\over 2}\theta^{\beta}\bar{\theta}^{\dot{\beta}}
\partial_{\beta\dot{\beta}}\phi -{i\over 2}\theta^2
\bar{\theta}^{\dot{\beta}}\partial^{\alpha}_{\dot{\beta}}
\psi_{\alpha}+{1\over 4}\theta^2 \bar{\theta}^2 \Box\phi.
$$

The known identities
\bea
& & \nabla^2\bar{\nabla}^2 + \bar{\nabla}^2\nabla^2
- \nabla\bar{\nabla}^2\nabla =
\Box_{+}=\Box-i\bar{W}^{\dot{\alpha}}\bar{\nabla}_{\dot{\alpha}} -
{i \over 2}(\bar{\nabla}\bar{W}),\nonumber \\
& & \bar{\nabla}^2\nabla^2 + \nabla^2\bar{\nabla}^2 -
\bar{\nabla}\nabla^2\bar{\nabla} =\Box_{-}=
\Box-iW^{\alpha}\nabla_{\alpha}-{i \over 2}(\nabla W)\nonumber
\eea
are also exploited.

\end{document}